\documentclass[sigconf]{acmart}

\usepackage{algorithmic}  
\usepackage{algorithm} 
\usepackage[algo2e]{algorithm2e} 
\usepackage{pifont,dsfont}
\usepackage{makecell}
\usepackage{amsfonts}
\usepackage{wrapfig,lipsum,booktabs}
\usepackage[export]{adjustbox}
\usepackage{xcolor,colortbl}
\usepackage{multirow}

\usepackage{graphicx}
\usepackage{subfigure}

\usepackage{refcount}
\usepackage{amsfonts}
\usepackage{caption}

\newcommand{\hide}[1]{}
\usepackage{xcolor,colortbl}

\newtheorem*{pro-stat}{Problem Definition}

\newcommand{\myModel}{{\textsc{MixRAG}}}

\AtBeginDocument{%
  \providecommand\BibTeX{{%
    \normalfont B\kern-0.5em{\scshape i\kern-0.25em b}\kern-0.8em\TeX}}}

\copyrightyear{2026}
\acmYear{2026}
\setcopyright{cc}
\setcctype{by}
\acmConference[WWW '26]{Proceedings of the ACM Web Conference 2026}{April 13--17, 2026}{Dubai, United Arab Emirates}
\acmBooktitle{Proceedings of the ACM Web Conference 2026 (WWW '26), April 13--17, 2026, Dubai, United Arab Emirates}
\acmPrice{}
\acmDOI{10.1145/3774904.3792680}
\acmISBN{979-8-4007-2307-0/2026/04}

\begin{document}

\title[MixRAG: Mixture-of-Experts Retrieval-Augmented Generation for Textual Graph Understanding \\ and Question Answering]{MixRAG : Mixture-of-Experts Retrieval-Augmented Generation for Textual Graph Understanding and Question Answering}







\author{Lihui Liu } 
\authornote{Lihui Liu is the first author and the corresponding author. Jiayuan Ding, Subhabrata Mukherjee and Carl Yang are collaborators.}
\email{hw6926@wayne.edu}
\affiliation{%
  \institution{Wayne State University}
  \city{Detroit}
  \state{Michigan}
  \country{USA}
}

\author{Jiayuan Ding}
\email{jiayuan@hippocraticai.com}
\affiliation{%
  \institution{Hippocratic AI}
  \city{Palo Alto}
  \state{California}
  \country{USA}
}

\author{Subhabrata Mukherjee}
\email{subho@hippocraticai.com}
\affiliation{%
  \institution{Hippocratic AI}
  \city{Palo Alto}
  \state{California}
  \country{USA}
}

\author{Carl Yang}
\email{j.carlyang@emory.edu}
\affiliation{%
  \institution{Emory University}
  \city{Atlanta}
  \state{Georgia}
  \country{USA}
}

\begin{abstract}

Large Language Models have achieved impressive performance across a wide range of applications. However, they often suffer from hallucinations in knowledge-intensive domains due to their reliance on static pretraining corpora. To address this limitation, Retrieval-Augmented Generation (RAG) enhances LLMs by incorporating external knowledge sources during inference. Among these sources, Textural Graphs offer structured and semantically rich information that supports more precise and interpretable reasoning. This has led to growing interest in Graph-based RAG systems. Despite their potential, most existing approaches rely on a single retriever to identify relevant subgraphs, which limits their ability to capture the diverse aspects of complex queries. Moreover, these systems often struggle to accurately judge the relevance of retrieved content, making them prone to distraction by irrelevant noise. To address these challenges, in this paper, we propose \myModel, a Mixture-of-Experts Graph-RAG framework that introduces multiple specialized graph retrievers and a dynamic routing controller to better handle diverse query intents. Each retriever is trained to focus on a specific aspect of graph semantics, such as entities, relations, or subgraph topology. A Mixture-of-Experts module adaptively selects and fuses relevant retrievers based on the input query. To reduce noise in the retrieved information, we introduce a query-aware GraphEncoder that carefully analyzes relationships within the retrieved subgraphs, helping to highlight the most relevant parts while down-weighting unnecessary noise. Empirical results show that our method achieves state-of-the-art performance and consistently outperforms various baselines.  The code can be found from \url{https://github.com/lihuiliullh/MixRAG}.

\end{abstract}




\maketitle

\section{Introduction}

Large Language Models have greatly advanced natural language processing, showing strong performance on many reasoning and text generation tasks~\cite{gpt2, google_lamda}.
However, they often struggle in knowledge-heavy areas, where accurate answers depend on up-to-date and detailed information. This is mainly because LLMs are trained on fixed datasets, which may be outdated, incomplete, or lack the specific knowledge needed for complex reasoning.
To address this shortcoming, Retrieval-Augmented Generation (RAG)~\cite{graphrag, replug, rag} has emerged as a prominent paradigm. By retrieving relevant context from external sources and conditioning LLM responses on the retrieved content, RAG enhances factuality, reduces hallucination, and supports knowledge-intensive tasks. 
While early RAG methods ~\cite{replug,rag} focused on retrieving unstructured documents (e.g., Wikipedia passages), recent efforts have explored retrieval over structured sources like textural graphs ~\cite{graphrag, gretriever}. These structured RAG systems aim to leverage the explicit semantics and relational structure of textural graphs to enable more grounded reasoning.

\begin{figure}[]
    \centering
    \includegraphics[width=0.4\textwidth]{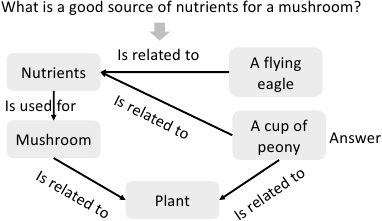} 
    \caption{An example of a retrieved subgraph for question answering. The subgraph contains both correct answer and noise data.}
    \label{example} 
\end{figure}

Despite promising progress, existing graph-based RAG methods face two key challenges that limit their effectiveness in real-world applications. First, most systems rely on a single retriever model to handle diverse query intents. However, textural graphs inherently encode multi-aspect information. For example, nodes often represent entities with domain-specific semantics (e.g., ``Vitamin D,'' ``Mushroom''), while edges express relational dependencies (e.g., ``part\_of,’’ ``requires’’). Using a single retriever trained on shallow lexical matching or embedding similarity often fails to capture this heterogeneity. Queries seeking complex relational inference may require different retrieval patterns than those involving single hop factual recall of entities. A one-size-fits-all retriever lacks the specialization to disentangle these nuances.

Second, retrieved subgraphs frequently contain irrelevant or noisy information, which can mislead the LLM during generation. Figure~\ref{example} shows a motivating case: when asked “What is a good source of nutrients for a mushroom?”, a standard retriever surfaces a subgraph containing entities such as “a flying eagle” and “a cut peony,” which are semantically disconnected from the query. Injecting such noisy knowledge into the LLM can degrade factual precision. This highlights a fundamental challenge: \textit{not all retrieved knowledge is equally useful}. Therefore, it raises an important question: \textit{How can we identify which retrieved knowledge is valuable and which is not?} This question becomes especially crucial when a significant portion of the retrieved subgraph lacks direct relevance or even introduces contradictions.

To address these limitations, we propose \myModel, a {Mixture-of-Expertsn~\cite{jiang2024mixtralexperts} Graph-RAG} framework. Instead of relying on a single retriever, our method introduces multiple expert retrievers, each trained to capture a specific aspect of graph semantics (e.g., entity names, relational paths, and local graph neighborhoods). A Mixture-of-Experts (MoE) controller is then used to dynamically combine suitable expert(s), based on query intent and expert specialization. This design enables flexible and query-adaptive retrieval that better aligns with the semantic demands of different question types.
Furthermore, to mitigate the noise in the retrieved subgraph, we design a query-aware \textit{GraphEncoder} module. This encoder performs fine-grained relational modeling over the retrieved graph fragments, learning to amplify useful signals and suppress distracting ones. The resulting graph-aware embeddings are used to augment the prompt space of the LLM, enabling more grounded and context-sensitive reasoning.
We conduct extensive experiments on the GraphQA benchmark and show that our method significantly outperforms strong baselines and achieves the state-of-the-art performance. Ablation studies further validate the contribution of expert-based retrieval and fine-grained subgraph embedding to downstream performance. In summary, we make the following contributions. 
\begin{itemize}
\item We introduce \textbf{Mixture-of-Experts Graph-RAG}, a novel RAG architecture that uses multiple specialized graph retrievers and a learned routing controller to merge relevant experts dynamically.
    
\item We propose a query-specific \textit{GraphEncoder} that encodes retrieved subgraphs based on structural and relational relevance, producing more informative embeddings for generation.

\item We empirically demonstrate state-of-the-art performance on the GraphQA benchmark, outperforming existing approaches. These results underscore the effectiveness of our proposed MOE-GraphRAG framework in capturing and leveraging multi-aspect knowledge for graph-based question answering.
\end{itemize}

The remainder of this paper is structured as follows. Section~\ref{problem-definition} defines the problem and outlines the notations adopted throughout the paper. Section~\ref{overview} presents the proposed framework along with its key technical components. Experimental results are discussed in Section~\ref{experiments}, while Section~\ref{related-work} reviews relevant literature. Finally, Section~\ref{conclusion} summarizes the main findings and concludes the paper.

\section{Problem Definition}\label{problem-definition}

\begin{table}[]
\centering
\caption{Main notations used in this paper.}
\small
\setlength\tabcolsep{3pt}
\begin{tabular}{|c|c|}
\hline
{\bf Symbol}       & {\bf Definition}                \\ \hline
$\mathcal{G}=(\mathcal{V}, \mathcal{R}, \mathcal{T})$ & A textual graph \\
$v_i$         & The $i^\textrm{th}$ text node (entity, sentence, or paragraph) in $\mathcal{G}$ \\
$r_i$         & The $i^\textrm{th}$ relation/edge in $\mathcal{G}$ \\
$q$ & The natural language query \\
$s_i$ & The $i$-th subgraph retrieved \\
$A_q$         & Answer set of $q$ \\
$a_q$           & A candidate answer of $q$ \\ 
$h_i$ & Dense embedding of node $v_i$ or relation $r_i$ in $\mathbb{R}^d$ \\ 
$h_q$ & Dense embedding of query $q$ in $\mathbb{R}^d$ \\ \hline
\end{tabular}
\label{notation}
\end{table}

Table ~\ref{notation} gives the main notations used throughout this paper. 
A textual graph is defined as $\mathcal{G}=(\mathcal{V}, \mathcal{R}, \mathcal{T})$, where $\mathcal{V}$ is the set of text nodes, $\mathcal{R}$ the set of relations, and $\mathcal{T}$ the set of factual triples. Unlike conventional knowledge graphs, nodes in textual graphs are free-form text units such as entities, sentences, or short paragraphs. Each triple is represented as $(h, r, t)$, where $h \in \mathcal{V}$ is the head node, $t \in \mathcal{V}$ the tail node, and $r \in \mathcal{R}$ the relation linking them.
Each node $v_i \in \mathcal{V}$ and relation $r_i \in \mathcal{R}$ can be encoded into a dense vector $h_i \in \mathbb{R}^d$. Similarly, a query $q$ is encoded into an embedding $h_q \in \mathbb{R}^d$.

\begin{figure}[t]
\centering
\includegraphics[width=0.48\textwidth]{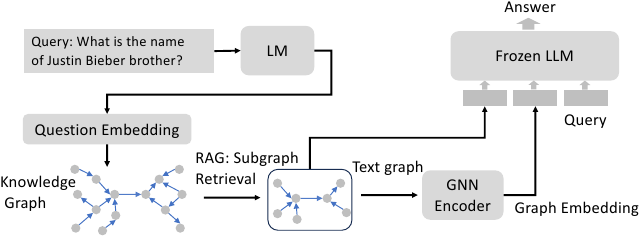}
\caption{Traditional retrieval-augmented generation (RAG) pipeline over textual graphs.}
\label{example2}
\end{figure}

\paragraph{Textual Graph Question Answering}
Textual Graph Question Answering (TGQA) aims to answer natural language questions by using information stored in a textual graph. The main challenge is to map an input question into a reasoning process over the graph to identify the correct answer node(s). Various approaches have been proposed, including few-shot prompting~\cite{wei2023chainofthoughtpromptingelicitsreasoning} and instruction tuning~\cite{gpt2}.
Recent work leverages Retrieval-Augmented Generation (RAG) to improve TGQA. Instead of directly predicting the answer, RAG first retrieves a relevant subgraph from the textual graph and then uses a large language model (LLM) to generate the answer based on this subgraph.
Specifically, given a question $q$, the retriever selects a subgraph $S \subseteq \mathcal{G}$ that contains relevant nodes and edges. The retrieved subgraph is then converted into a textual format—such as a sequence of triples or natural language descriptions—which is combined with the original question to form the input to the LLM. The LLM uses this augmented input to generate the answer $a$.
Alternatively, the retrieved subgraph can be encoded with a graph neural network (GNN)~\cite{kipf2016semi,hamilton2017inductive} to produce embeddings that capture both semantic and structural information. The LLM can attend to both the textualized subgraph and the GNN embeddings, resulting in more accurate and grounded reasoning (Figure~\ref{example2}).

Mixture-of-Experts (MoE) was originally proposed in \cite{shazeer2017outrageouslylargeneuralnetworks} as a strategy to improve model expressiveness by combining multiple specialized “expert” networks, with a gating mechanism dynamically selecting which experts to use for each input. This approach is more powerful than a single monolithic model, as it allows the system to adaptively leverage the most relevant expertise for each instance, enabling better representation, higher capacity, and improved generalization across diverse inputs.

\paragraph{Limitations and Motivation for Mixture-of-Experts}
Standard RAG-based methods typically rely on a single retrieval perspective (e.g., entity-level or relation-level), which may overlook complementary evidence. To address this, we propose a Mixture-of-Experts (MoE) retrieval framework. Each expert focuses on a distinct retrieval signal—such as entity-centric, relation-centric, or subgraph-based retrieval—and the system dynamically selects among experts. This enables the model to gather multi-perspective evidence from the textual graph, which is then used to guide the LLM in generating faithful and accurate answers.

Formally, the problem this paper studies is defined as follows. 
\begin{pro-stat}{Answering Query TGQA: }

	\textbf{Given:} (1) A textual graph $\mathcal{G}$, (2) an  natural language question; 
	
	\textbf{Output:} The answer of the question.
\end{pro-stat}

\section{Proposed Method}\label{overview}


In this section, we introduce the details of the proposal method, we begin with multiaspect knowledge retrieval. We then delve into two specific
modules: semantic reasoning and subgraph reasoning, providing detailed explanations of their functionalities, and finally, we introduce how to fuse them together. 

The key idea behind our model is that real-world questions vary in complexity. Some questions are simple and can be easily answered, in which case a naive entity retriever is sufficient to find the correct answer. Using a more complex subgraph retriever for such questions may introduce unnecessary noise and even degrade performance. On the other hand, more complex questions require multi-step reasoning, where a sophisticated retriever—such as a subgraph-based approach—can provide richer context and significantly improve the model’s reasoning capabilities.
By adopting different retrievers for different types of questions and then combining their outputs, the system can adapt to the complexity of each question and achieve better overall performance.

\subsection{Retrieval of Multi-Aspect Knowledge}

Previous TGQA methods typically focus on retrieving either entities and relations~\cite{Gu_2021} or multiple triples to construct subgraphs~\cite{gretriever}. In contrast, we propose to simultaneously leverage {three complementary types of knowledge}: entities, relations, and subgraphs. 
Each captures a different retrieval mechanism that is suitable for different types of queries.
By jointly utilizing different knowledge types, we enable more robust and accurate alignment between questions and the corresponding knowledge components. Here, we describe each retrieval in detail.

\textbf{Entity Retrieval.}
Entity retrieval selects a small set of candidate entities from the graph that are most relevant to the input question. It directly predicts which entities are likely to be the answer or closely related to the question based on their similarity in the embedding space.
The question is encoded into a dense vector using a language model and compared against entity embeddings. The top-$k$ most similar entities are selected during the reasoning process. 
The motivation behind this retriever is that some questions can be answered with minimal reasoning; in such cases, a simple entity retriever may outperform more complex subgraph-based retrieval methods, which can introduce unnecessary noise.

\begin{figure*}[h!]
    \centering
    \includegraphics[width=0.95\textwidth]{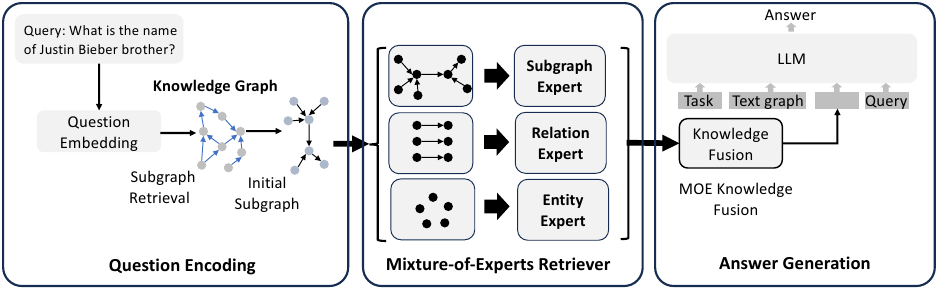} 
    \caption{The framework of \myModel.}
    \label{example} 
\end{figure*}

\textbf{Relation Retrieval.}
Complementing entity retrieval, relation retrieval identifies the most relevant triples that reflect the semantic intent of the question. It predicts which triples are likely to be the answer or closely related to the question.
Like entities, relations are encoded as embeddings using their names and descriptions. The top-$k$ matching triples are then selected based on their similarity to the question embedding.
The idea behind a relation retriever is similar to that of an entity retriever, but it leverages richer semantic information. Instead of focusing solely on individual entities, it also considers the relations between pairs of entities, capturing more contextual information and providing a more comprehensive understanding compared with a simple entity retriever.

\textbf{Subgraph Retrieval.}
Subgraph retrieval aims to extract a compact yet comprehensive subgraph that is most relevant to the question. It is designed to capture richer information needed for answering complex queries.
It serves two main goals: filtering out irrelevant information that could distract the language model from the essential context, and maintaining a manageable graph size that can be effectively serialized into text for LLM processing.
The subgraph is typically constructed by expanding from the retrieved seed entities along top-ranked relations, within a limited number of hops. This process ensures that the resulting subgraph preserves both the structural and semantic connections most relevant to the query.
Unlike entity and relation retrievers, a subgraph retriever extracts richer information from the underlying data graph, enabling support for more complex reasoning tasks.

\subsection{Semantic Reasoning with Entity and Relation Retrieval}

In the previous subsection, we briefly introduced the three retrievers and their respective roles. Here, we focus on the details of Entity Retrieval and Relation Retrieval, which together form the basis of Semantic Reasoning. Semantic reasoning aims to identify the components of a textual graph that are most semantically aligned with the input query. Within this framework, entity and relation retrieval act as complementary processes: entity retrieval predicts the most likely answer entities, while relation retrieval identifies the most relevant triples.

Formally, we represent the initial subgraph retrieved by subgraph retriever as \( G = (E, R, T) \), where \( E \) denotes the set of entities, \( R \) the set of relations, and \( T \subseteq E \times R \times E \) the set of textualized triples. Each entity \( e_i \in E \) is encoded into a dense vector \( h_i \in \mathbb{R}^d \). For each triple \( (h, r, t) \in T \), we construct an initial representation by concatenating the head, relation, and tail embeddings into a vector \( [h_h ; h_r ; h_t] \in \mathbb{R}^{3d} \), which is subsequently projected through a learned transformation \( W_t \in \mathbb{R}^{d \times 3d} \). This ensures that both entity-level and triple-level representations reside in a unified semantic space.
Given a natural-language query \( q \), its embedding \( h_q \in \mathbb{R}^d \) is obtained using an instruction-tuned language model. To measure the relevance of each graph element to the query, we concatenate its representation with the query embedding, $z_i = [\,h_i \,\|\, h_q\,]$, 
and compute a relevance logit using a learnable scoring function \( f(\cdot) \), $\phi_i = f(z_i)$.
This parameterized compatibility function enables the model to learn query-specific semantic matching over entities or triples.

To perform differentiable selection over graph elements, we adopt the Gumbel-Softmax continuous relaxation. Using the reparameterization trick, we perturb the logits with Gumbel noise and compute
\[
p_i = \mathrm{softmax}\!\left( \frac{\phi_i + g_i}{\tau} \right),
\qquad g_i \sim \mathrm{Gumbel}(0,1),
\]
where \( \tau \) is the temperature controlling the sharpness of the distribution and is gradually annealed during training. The resulting distribution \( p \in \mathbb{R}^{|E|} \) (for entities) or \( p \in \mathbb{R}^{|T|} \) (for triples) represents attention weights over the textual graph, highlighting the most relevant components with respect to the query.

Finally, we compute a query-specific graph representation by first selecting the top nodes or edges using the differentiable top-\(k\) sampler, and then encoding the resulting masked subgraph with a GNN. 
Using these GNN-refined embeddings, we can optionally pool over the selected nodes to obtain a graph-level representation
\[
\tilde{h}_G = \operatorname{POOL}(\{h_i \mid m_i = 1\}),
\]
which yields a compact, semantically aligned summary of the most relevant entities or triples with respect to the input query. This query-specific representation is subsequently used by downstream reasoning modules to generate contextually grounded answers.

\subsection{Subgraph Retreiver and Reasoning}

While entity and relation retrieval focus on pinpointing specific knowledge components, subgraph retrieval aims to extract a structured and compact set of interconnected facts that collectively support more complex reasoning. 
In this paper, we follow the idea of GRetreiver framework~\cite{gretriever} to identify a relevant subgraph conditioned on the input query $q$. 
Let \( G = (V, E) \) denote the textual graph, where both nodes and edges are associated with textual descriptions.
We encode the query and the graph components using a pretrained language model, Sentence-BERT~\cite{sentencebert}. 
{\small
\begin{align}
\mathbf{z}_q &= \mathrm{LM}(q), \quad
\mathbf{z}_{v_i} = \mathrm{LM}(v_i), \quad
\mathbf{z}_{e_{i,j}} = \mathrm{LM}(e_{i,j})
\end{align}
}

We then compute cosine similarities between the query embedding \( \mathbf{z}_q \) and all node and edge embeddings, and retrieve the top-\( k \) nodes \( V_k \) and edges \( E_k \) with the highest similarity scores. To ensure the subgraph is connected and informative, we apply the {Prize-Collecting Steiner Tree (PCST)} algorithm~\cite{gretriever}. Each retrieved node or edge is assigned a prize based on its similarity rank. The final subgraph \( S \subseteq G \) is selected to maximize the total prize while minimizing edge costs:

{\small
\begin{align*}
S = \underset{S \subseteq G}{\operatorname{argmax}} \left(
\sum_{v_i \in V_k} \text{prize}(v_i) +
\sum_{e_{i,j} \in E_k} \text{prize}(e_{i,j})
- c \cdot |E_S|
\right)
\end{align*}
}
This subgraph serves as both a textual context for the LLM and an input to graph encoders for downstream reasoning.

After retrieving the subgraph, the next step is to perform reasoning over it. 
Prior methods such as G-Retriever employ GCNs~\cite{gcn} or GATs~\cite{gat} to encode subgraphs. However, these architectures often suffer from the over-smoothing problem~\cite{oversmooth}, where node embeddings become indistinguishable after several layers of message passing. 
This issue is particularly problematic in our setting, where the retrieved subgraphs may include a mix of relevant and irrelevant (noisy) nodes. 
Therefore, it is crucial to generate \textit{query-aware} representations that emphasize the most relevant nodes and edges with respect to the input query.
For example, consider a query involving \texttt{a cut peony}. The encoder should highlight the node corresponding to \texttt{a cut peony} while downweighting unrelated nodes such as \texttt{a flying eagle}, even if they are structurally nearby in the graph. 
To achieve this, we design a query-conditioned GNN in which both message passing and node interactions are dynamically modulated by the input query \( q \).

Concretely, we redefine the attention weights over edges using a query-aware mechanism. At each GNN layer \( l \), the attention weight \( \zeta_{e_{i,j}}^{(l)} \) for edge \( e_{i,j} \) is computed as:
\begin{align}
\alpha_{v_i}^{(l)} &= \texttt{LINEAR}\left(\texttt{CONCAT}(z_{v_i}^{(l)}, q)\right) \\
\beta_{v_j}^{(l)} &= \texttt{LINEAR}\left(\texttt{CONCAT}(z_{v_j}^{(l)}, q)\right) \\
\gamma_{e_{i,j}} &= \texttt{LINEAR}\left(\texttt{CONCAT}(z_{e_{i,j}}, q)\right) \\
\zeta_{e_{i,j}}^{(l)} &= \texttt{tanh}\left(\alpha_{v_i}^{(l)} + \gamma_{e_{i,j}} - \beta_{v_j}^{(l)}\right)
\end{align}
Here, \( \alpha_{v_i}^{(l)} \), \( \beta_{v_j}^{(l)} \), and \( \gamma_{e_{i,j}} \) are intermediate representations that encode query-conditioned information for the source node, target node, and edge respectively. The resulting attention weight \( \zeta_{e_{i,j}}^{(l)} \) modulates how much influence the message passed along edge \( e_{i,j} \) should have.

The message from node \( v_i \) to node \( v_j \) at layer \( l \) is generated as:
\begin{equation}
\text{msg}_{e_{i,j}}^{(l)} = \texttt{LINEAR}\left(\texttt{CONCAT}(z_{v_i}^{(l)}, z_{v_j}^{(l)}, z_{e_{i,j}}, q)\right)
\end{equation}
Each node then updates its representation using the attention-weighted aggregation of incoming messages:
\begin{equation}
z_{v_j}^{(l+1)} = \frac{1}{d_{v_j}}\sum_{v_i \in \mathcal{N}(v_j)} \zeta_{e_{i,j}}^{(l)} \cdot \text{msg}_{e_{i,j}}^{(l)}
\end{equation}

Unlike standard GCNs that apply fixed, query-independent filters, our query-conditioned GNN dynamically adapts both message generation and propagation to the specific information needs encoded in the query \( q \). This enables the model to attend to semantically aligned regions of the subgraph while filtering out unrelated content.
After \( L \) layers of message passing, we obtain the final node representations \( z_{v_j}^{(L)} \) for each node \( v_j \) in the subgraph \( S \). 

\subsection{Mixture of Experts Gating}

In the previous subsection, we introduced various types of information retrievers. Here, we describe how to effectively fuse the retrieved signals. To accommodate the diverse reasoning requirements in textual graphs, we adopt a Mixture-of-Experts (MoE) framework that dynamically integrates knowledge from multiple specialized, query-aware reasoning modules: the entity (node) expert \(f_e\), the relation (edge) expert \(f_r\), and the subgraph expert \(f_s\). This design allows the model to adaptively leverage the most relevant processing pathway for each node in the graph.

Let \(o_v^{(i)}\in\mathbb{R}^d\) denote the node-level output produced by expert \(f_i\in\{f_e,f_r,f_s\}\) for node \(v\). Instead of computing a single global gating score, we use a node-wise, graph-aware gating network. For each node, we construct a gate input by combining its original node feature with features aggregated from its local graph context with gnn. This input is fed into a small MLP to produce per-expert logits,  
\[
\boldsymbol{\phi}_v \;=\; \mathrm{MLP}(\text{node feature + learned contextual features}) \in \mathbb{R}^{3}.
\]  
These logits are converted into a node-wise probability distribution over experts using a softmax function:  
\[
\alpha_{v,i} \;=\; \frac{\exp(\phi_{v,i})}{\sum_{j\in\{e,r,s\}}\exp(\phi_{v,j})},\qquad i\in\{e,r,s\},
\]  
so that \(\sum_i \alpha_{v,i} = 1\) for each node \(v\).

The final fused representation at node \(v\) is the convex combination of expert outputs weighted by the node-wise gate,  
\[
\tilde{\mathbf{h}}_v \;=\; \sum_{i\in\{e,r,s\}} \alpha_{v,i}\, o_v^{(i)}.
\]  
When a graph-level soft prompt is required, the node-wise fused embeddings are pooled over the selected nodes to form a compact graph representation:  
\[
\mathbf{p}_{\text{soft}} \;=\; \operatorname{MeanPool}\big(\{\tilde{\mathbf{h}}_v\}\big),
\]  
which serves as a query-conditioned graph prompt for downstream modules.

\subsection{Answer Generation}

After obtaining the fused information from multiple retrievers, we prepare the input for response generation using a large language model (LLM). The goal is to generate an answer that is grounded in both the user's query \( q \) and the retrieved supporting evidence.
To incorporate task-specific control signals, we prepend a trainable soft prompt \( p_{\text{soft}} \) to the input. This soft prompt is produced by a Mixture-of-Experts (MoE) controller that dynamically selects and combines expert embeddings based on the current input context as described in the last subsection. It serves as a continuous prefix that guides the LLM toward producing task-aware and knowledge-grounded responses. The retrieved subgraph \( G_{\text{sub}} \) is verbalized into natural language using predefined templates that describe entities and relations in sentence form. 

The complete input to the LLM consists of four parts: the soft prompt \( p_{\text{soft}} \), the tokenized task instruction \( p_{\text{task}} \), the textualized subgraph \( p_{\text{text}} \), and the user query \( q \). 
\begin{equation}
a_{\text{gen}} = \texttt{LLM}([ p_{\text{task}}; p_{\text{soft}}; p_{\text{text}}; q]),
\end{equation}

Here, \( a_{\text{gen}} \) is the final response generated by the LLM. By integrating multi-aspect information, the model is guided to produce coherent, grounded, and contextually relevant responses. 

\subsection{Theoretical Analysis}

In this section, we provide a theoretical justification that our proposed framework subsumes existing graph-based retrieval-augmented generation (RAG) methods as special cases. The central idea is that our model integrates three complementary retrievers—entity-, relation-, and subgraph-based—within a Mixture-of-Experts (MoE) architecture, whereas most prior work relies on a single retriever, typically subgraph-based.

Let $\mathcal{F}_{\text{ours}}$ denote the hypothesis class induced by our framework. Given a query $q$, each expert $f_i \in \{f_e, f_r, f_s\}$ produces node-level, query-conditioned representations over the graph. A node-wise MoE controller computes routing weights $\alpha_{v,i}(q)$ for each node $v$ and expert $i$, satisfying
\[
\sum_{i \in \{e,r,s\}} \alpha_{v,i}(q) = 1,
\qquad
\alpha_{v,i}(q) \geq 0.
\]
Here, the gating weights are produced by a learnable controller that depends on the query and aggregated graph context.

Now consider an existing graph RAG method with hypothesis class $\mathcal{F}_{\text{base}}$ that employs only a single retriever, such as a subgraph retriever $f_s$. Such a method can be realized as a special case of our framework by choosing the node-wise routing weights as
\[
\alpha_{v,s}(q) = 1,
\qquad
\alpha_{v,e}(q) = \alpha_{v,r}(q) = 0,
\quad \forall v \in \mathcal{V}.
\]
Under this setting, the node-level fusion reduces to
\[
\tilde{\mathbf{h}}_v(q) = f_s(q)_v,
\]
which exactly matches the graph representation used in prior subgraph-based RAG approaches~\cite{gretriever}. Analogously, enforcing $\alpha_{v,e}(q)=1$ or $\alpha_{v,r}(q)=1$ for all nodes recovers pure entity-based or relation-based retrieval methods, respectively. Thus, existing methods correspond to specific points in the space of routing weights defined by our MoE controller.

\paragraph{Implication.}
This analysis establishes that our framework is strictly more general than existing graph RAG approaches. Since our MoE controller can always collapse to a single expert, we can guarantee that our method performs at least as well as existing baselines in the worst case. More importantly, by dynamically combining multiple retrieval perspectives, our framework has the potential to improve performance on complex queries where complementary evidence from entities, relations, and subgraphs is required. 
\label{sec:method}

\section{Experiments}\label{experiments}

\begin{table*}[ht]
\centering
\caption{Statistics of datasets.}
\label{tab:dataset_statistics}
\begin{tabular}{|c|c|c|c|c|}
\hline
\textbf{Dataset} & \textbf{ExplaGraphs} & \textbf{SceneGraphs} & \textbf{WebQSP}  \\ \hline
\#Graphs          & 2,766               & 100,000             & 4,737                    \\ \hline
Average \#Nodes   & 5.17                & 19.13               & 1370.89                  \\ \hline
Average \#Edges   & 4.25                & 68.44               & 4252.37                 \\ \hline
Node Attribute    & Commonsense concepts & Object attributes    & Entities in Freebase  \\ \hline
Edge Attribute    & Commonsense relations & Spatial relations   & Relations in Freebase  \\ \hline
Task              & Commonsense reasoning & Scene graph QA       & KGQA             \\ \hline
\end{tabular}
\label{dataset}
\end{table*}

\begin{table*}[ht]
\centering
\caption{Performance comparison across ExplaGraphs, SceneGraphs, and WebQSP datasets for different configurations, including Inference-only, Frozen LLM with prompt tuning (PT), and Tuned LLM settings. Mean scores and standard deviations (mean ± std) are presented. The best result for each task is highlighted in \textbf{bold}, and the second best result is \underline{underlined}.}
\begin{tabular}{l l c c c}
\toprule
\textbf{Setting} & \textbf{Method} & \textbf{ExplaGraphs} & \textbf{SceneGraphs} & \textbf{WebQSP} \\
\midrule
\multirow{4}{*}{Inference-only}
& Zero-shot & 0.5650 & 0.3974 & 41.06 \\
& Zero-CoT~\cite{kojima2023largelanguagemodelszeroshot} & 0.5704 & 0.5260 & 51.30 \\
& CoT-BAG~\cite{wang2024languagemodelssolvegraph} & 0.5794 & 0.5680 & 39.60 \\
& KAPING~\cite{baek-etal-2023-knowledge} & {0.6227} & 0.4375 & 52.64 \\
\midrule
\multirow{3}{*}{Frozen LLM w/ PT}
& Prompt tuning & 0.5763 ± 0.0243 & 0.6341 ± 0.0024 & 48.34 ± 0.64 \\
& GraphToken~\cite{perozzi2024letgraphtalkingencoding} & 0.8508 ± 0.0551 & 0.4903 ± 0.0105 & 57.05 ± 0.74 \\
& G-Retriever & {0.8516} ± 0.0092 & {0.8131} ± 0.0162 & {70.49} ± 1.21 \\
\midrule
\multirow{2}{*}{Tuned LLM}
& LoRA & {0.8538} ± 0.0353 & 0.7862 ± 0.0031 & 66.03 ± 0.47 \\
& G-Retriever w/ LoRA & \underline{0.8705 ± 0.0329} & \underline{0.8683 ± 0.0072} & \underline{73.79 ± 0.70} \\
& \myModel\ w/ LoRA & \textbf{0.8863 ± 0.0288}  & \textbf{0.8712 ± 0.0064} & \textbf{75.31  ± 0.81} \\
\bottomrule
\end{tabular}
\label{tab:performance}
\end{table*}

In this section, we present the experimental results of \myModel, demonstrating its effectiveness and flexibility in graph-based reasoning tasks. Our experiments aim to validate the benefits of combining multiple retrievers in a Mixture-of-Experts framework, including entity, relation, and subgraph retrievers, and to analyze how each component contributes to overall performance across different datasets and query types. We also examine hyperparameters such as the number of GraphEncoder layers and visualize how the model distributes attention among experts to adapt to varying reasoning requirements.

\subsection{Experimental Setup}
\noindent\textbf{Datasets.}  
We evaluate on the GraphQA benchmark~\cite{gretriever}, which includes three datasets: ExplaGraphs, SceneGraphs, and WebQSP. These datasets span a diverse range of reasoning requirements over textural graphs. ExplaGraphs is a dataset for generative commonsense reasoning, SceneGraphs targets spatial reasoning over visual scene graphs, and WebQSP involves complex natural language questions over Freebase-derived subgraphs. The statistics for each dataset are summarized in Table~\ref{dataset}. 

\noindent\textbf{Metrics.}  
Following GraphQA~\cite{gretriever}, we use {accuracy} (ACC) for both ExplaGraphs and SceneGraphs, as they are treated as single-answer classification tasks. For WebQSP, which often has multiple valid answers per query, we report {Hit@1}, where a prediction is considered correct if it matches any ground truth answer. This allows for flexible evaluation under one-to-many supervision.

\noindent\textbf{Baselines.}  
We compare our method against two main categories of baselines. (1) {Inference-only methods}: these include Zero-shot prompting, Zero-CoT~\cite{kojima2023largelanguagemodelszeroshot}, CoT-BAG~\cite{wang2024languagemodelssolvegraph}, and KAPING~\cite{baek-etal-2023-knowledge}, which do not incorporate graph-structured knowledge or prompt adaptation. (2) {Prompt-tuning methods}: these include Prompt Tuning, GraphToken~\cite{perozzi2024letgraphtalkingencoding}, and the state-of-the-art G-Retriever~\cite{gretriever}, which leverage graph-derived prompts or neural retrievers to inject external knowledge into LLMs. For \myModel, we set $k$ as 20.




\subsection{Effectiveness of \myModel}

The results are summarized in Table~\ref{tab:performance}, which compares \myModel{} against all baseline methods. We report performance under varying model settings: inference-only, frozen LLM with prompt tuning, and tuned LLM. As we can see, among inference-only baselines, KAPING achieves the strongest results on datasets ExplaGraphs and WebQSP, outperforming Zero-CoT and CoT-BAG. While CoT-BAG achieves the best performance on SceneGraphs. However, once prompt tuning is enabled over a frozen LLM, performance improves substantially. For example, G-Retriever shows significant gains across all datasets, especially on SceneGraphs (0.8131) and WebQSP (70.49), outperforming GraphToken and basic prompt tuning by large margins. This highlights the advantage of retrieval-aware soft prompting in handling textual-graph reasoning tasks.

We see further performance improvements when allowing lightweight finetuning of the LLM. Using LoRA-based tuning~\cite{hu2021loralowrankadaptationlarge} leads to better results than keeping the LLM frozen. For example, it achieves 66.03 on WebQSP, compared to 57.05 from GraphToken. Interestingly, G-Retriever combined with LoRA consistently performs better than using LoRA alone, suggesting that structured retrieval and efficient tuning work well together. Finally, our method, \myModel\ with LoRA, achieves new state-of-the-art results on all three datasets: 0.8863 on ExplaGraphs, 0.8712 on SceneGraphs, and 75.31 on WebQSP. These results show that combining multi-aspect graph retrieval with a mixture-of-experts design helps the model generalize well across different reasoning tasks.

\subsection{Ablation Study}

\begin{table}[ht]
\centering
\caption{Accuracy of Different Expert Combinations. Results on SceneGraphs are omitted due to high computational cost.}
\begin{tabular}{lcc}
\toprule
\textbf{Expert Combination} & \textbf{ExplaGraphs} & \textbf{WebQSP} \\
\midrule
Only Entity           & 0.8247 & 67.76 \\
Only Relation         & 0.8466 & 69.89 \\
Only Subgraph         & 0.8765 & 73.99 \\
Entity + Relation     & 0.8574 & 71.92 \\
Entity + Subgraph     & 0.8646 & 72.91 \\
Relation + Subgraph   & 0.8682 & 73.03 \\
\myModel{} (All)      & \textbf{0.8863} & \textbf{75.31} \\
\bottomrule
\end{tabular}
\label{tab:performance}
\end{table}

\textbf{Retreiveral Fusion:}
We begin by evaluating the performance of different retrieval strategies. Specifically, we compare the effectiveness of using the entity retriever, the relation retriever, and the subgraph retriever. The results are summarized in Table~\ref{tab:performance}. 
As shown, using only the subgraph retriever leads to better performance than using either the entity or relation retriever alone, highlighting the importance of structural context. While combining entity and relation retrieval brings some improvement, combinations that include subgraph retrieval consistently perform better. Notably, our full model (\myModel) achieves the best results on both datasets, showing that integrating entity, relation, and subgraph information enables more complete and semantically grounded reasoning for textual graph question answering.

\begin{figure}[t]
    \centering
    \includegraphics[width=0.35\textwidth]{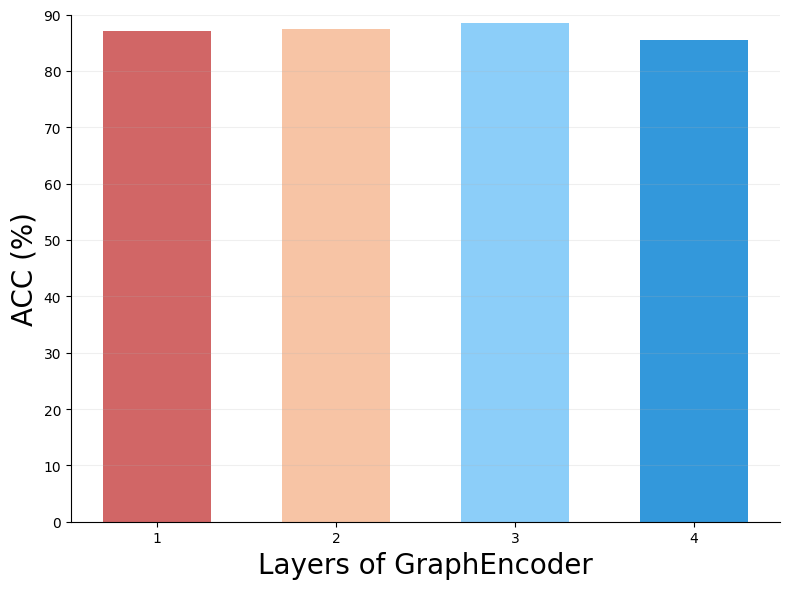} 
    \caption{Accuracy of \myModel{} on ExplaGraphs with respect to different numbers of GraphEncoder layers.}
    \label{layers} 
\end{figure}

\textbf{Number of GNN layers:}
In the hyperparameter study, we evaluate how the number of GraphEncoder layers influences model performance. As shown in Figure~\ref{layers}, accuracy improves consistently when increasing the number of layers from one to three, with three layers yielding the highest accuracy. However, adding a fourth layer results in performance degradation, likely due to oversmoothing. These results highlight the critical role of encoder depth in balancing expressive power and oversmoothing risk. They also suggest that deeper encoders do not necessarily translate to better performance in textual graph reasoning. Overall, using two or three layers is sufficient to capture the necessary structural and semantic information for effective question answering over textual graphs in \myModel{}.

\begin{figure}[t]
    \centering
\includegraphics[width=0.37\textwidth]{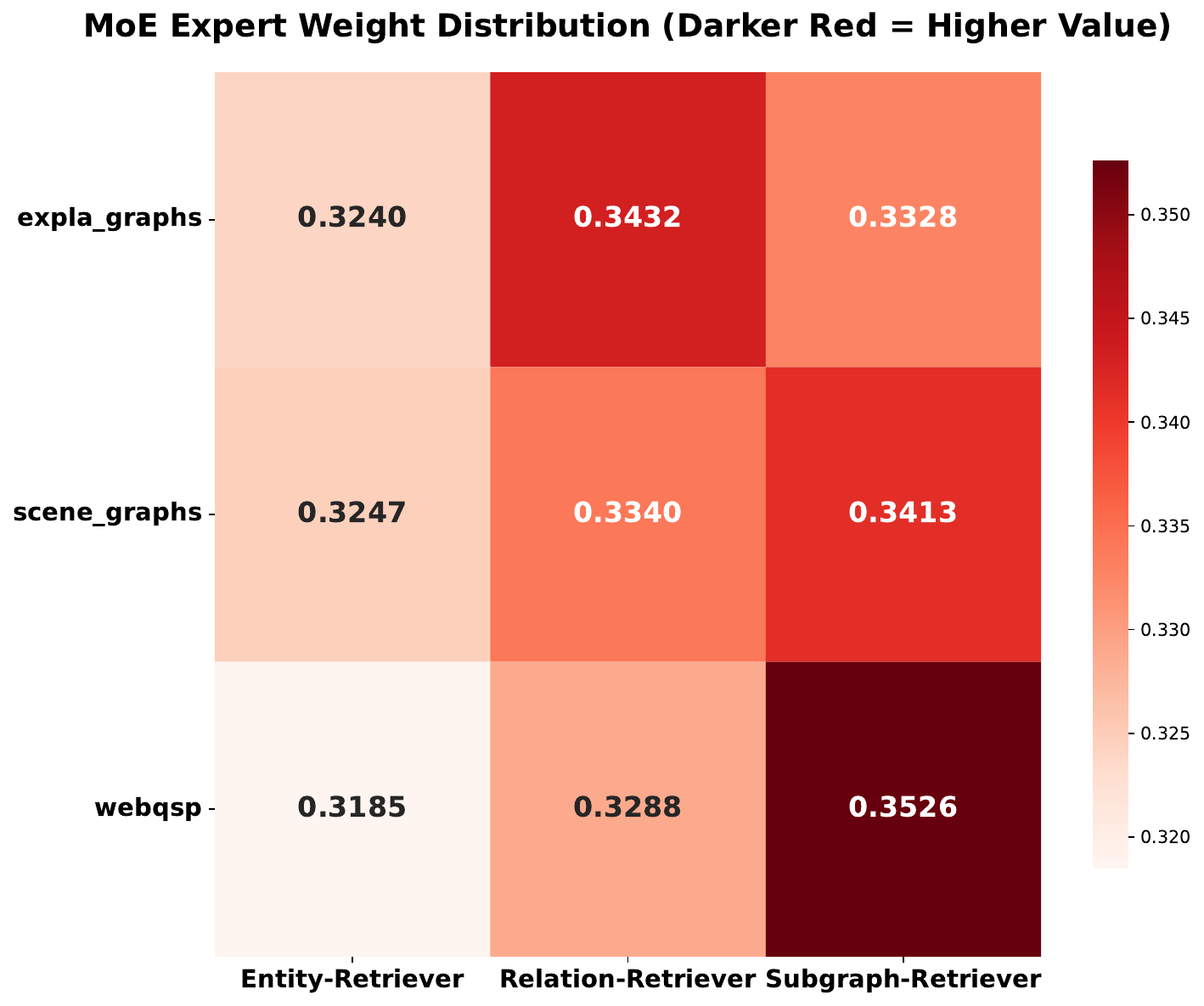} 
    \caption{Distribution of Expert Weights Across Different Datasets (Darker Red = Higher Value).}
    \label{heatmap} 
\end{figure}

\noindent\textbf{Distribution of Expert Weights Across Different Datasets:}
Figure~\ref{heatmap} shows how the model distributes attention across different experts for each dataset (we select a subset of data points). For ExplaGraphs, the relation retriever receives the highest weight, which makes sense because the task involves comparing relations between concepts to determine whether one claim supports or contradicts another. In contrast, for SceneGraphs and WebQSP, the subgraph retriever plays the most important role, followed by the relation retriever, while the entity retriever contributes the least. These patterns reflect how different datasets require different types of reasoning, and \myModel{} adapts its retrieval strategy to meet those needs.

\begin{figure}[t]
    \centering
\includegraphics[width=0.37\textwidth]{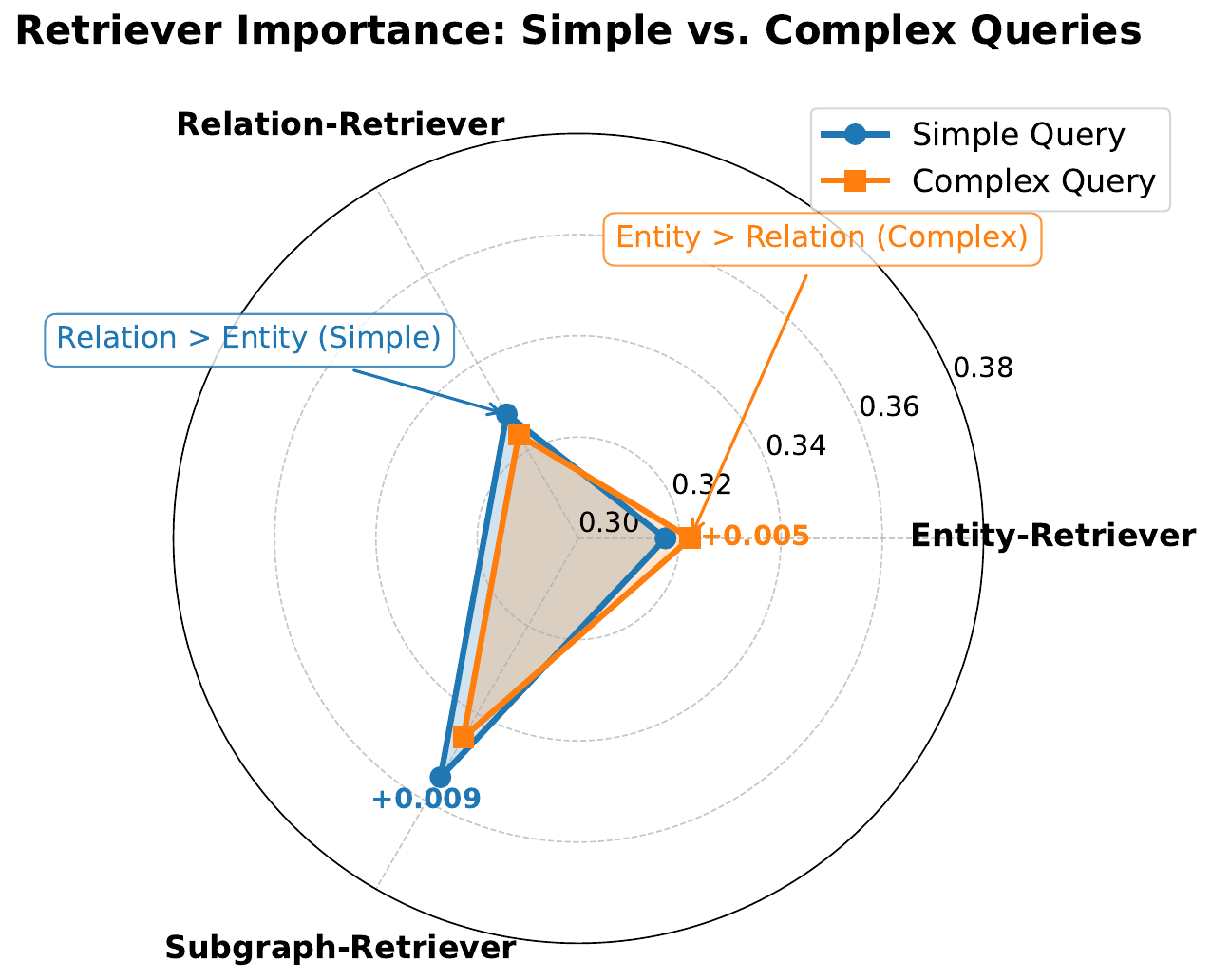} 
    \caption{Distribution of Expert Weights Across Different Query Types in WebQSP.}
    \label{radar} 
\end{figure}

\noindent\textbf{Distribution of Expert Weights Across Different Query Types in WebQSP:}
Figure~\ref{radar} shows how the model assigns expert weights to different retrievers for different types of queries in WebQSP. We divide the queries into two types: simple queries, which are 1-hop questions, and complex queries, which require multiple reasoning steps.
From the figure, we can see that the subgraph retriever always gets the highest weight, no matter if the query is simple or complex. The difference lies in how the relation and entity retrievers are used. For simple queries, the relation retriever is more important than the entity retriever. This makes sense because a 1-hop query often needs just one relation to find the answer. But for complex queries, the entity retriever becomes more important. This is likely because the relation retriever may struggle to connect multiple relations and can introduce noise. In such cases, directly retrieving relevant entities works better and improves accuracy.

\section{Related work}\label{related-work}

\paragraph{Retrieval-Augmented Generation (RAG)}  
RAG has emerged as a powerful paradigm for mitigating key limitations of large language models (LLMs), particularly their tendency to hallucinate or produce factually inaccurate responses in knowledge-intensive settings~\cite{gao-rag, sun2024survey,graphmcts,unifying_knowledge}. By conditioning generation on retrieved external knowledge, RAG improves factual grounding and task-specific adaptability.
Existing RAG approaches can be broadly categorized into three main paradigms. (1) Naive RAG adopts a straightforward pipeline consisting of indexing, retrieval, and generation, typically relying on embedding-based retrievers to identify relevant information~\cite{ma-rag}. (2) Advanced RAG enhances retrieval effectiveness through techniques applied before and after retrieval. Pre-retrieval methods include query rewriting, expansion, or transformation~\cite{peng-rewriting, zheng-rewriting}, while post-retrieval methods involve reranking the retrieved candidates based on their relevance~\cite{qin-rerank}. (3) Modular RAG provides greater flexibility by incorporating multiple types of data, such as unstructured text, structured tables, and knowledge graphs. It also utilizes large language models (LLMs) to generate or refine retrieval queries~\cite{yu2023generate}. These methods support more robust and adaptable retrieval strategies, making them especially suitable for handling complex reasoning tasks.

\paragraph{LLMs and Knowledge Graphs.}
Knowledge graph reasoning has been studied for a long time. ~\cite{binet, liu-2025-hyperkgr}.
Combining structured knowledge with large language models (LLMs) has been shown to improve factual accuracy, enhance interpretability, and boost reasoning performance. Broadly, there are three main strategies for integrating LLMs with knowledge graphs (KGs). First, KG-enhanced LLMs incorporate information from KGs either during pretraining~\cite{liu2020k, sun-etal-2020-colake} or at inference time through retrieval-based conditioning~\cite{lewis2020retrieval, wang-etal-2024-boosting-language}. Second, LLM-augmented KGs use LLMs to support tasks such as KG construction~\cite{bosselut-etal-2019-comet}, completion~\cite{kim-etal-2020-multi}, and representation learning~\cite{wang2023reasoningmemorizationnearestneighbor}. Third, synergistic integration refers to a co-evolution process in which the KG guides LLM inference ~\cite{liu2024logic}, while the LLM simultaneously enriches or restructures the KG. This mutual reinforcement leads to more powerful and flexible reasoning capabilities~\cite{yasunaga2022dragon}.

\paragraph{Mixture-of-Experts}  
The Mixture of Experts framework ~\cite{moe} has established itself as a fundamental paradigm in machine learning for developing adaptive systems. Initial work focused on traditional machine learning implementations ~\cite{NIPS1996_6c8dba7d}, with subsequent breakthroughs emerging through its integration with deep neural networks ~\cite{gcn}.
More recently, researchers have explored applying MoE approaches to in-context learning scenarios ~\cite{onepromptisnotenough}, demonstrating their potential to enhance large language model performance.

\section{Conclusion}\label{conclusion}

In this paper, we propose \myModel, a Mixture of Expert Retrieval-Augmented Generation framework that combines multiple specialized graph retrievers to better match query intent. Each retriever focuses on a different aspect of graph semantics, enabling more accurate and flexible retrieval. To reduce noise in the retrieved subgraphs, we introduce a query-aware GraphEncoder that highlights relevant information and filters out distractions. Experiments demonstrate that our approach outperforms strong baselines, achieving state-of-the-art performance.

\subsection{Acknowledge}

This research was partially supported by the US National Science Foundation under Award Number 2442172 and the US National Institute of Diabetes and Digestive and Kidney Diseases of the US National Institutes of Health under Award Number K25DK135913.

\bibliographystyle{ACM-Reference-Format}
\bibliography{008reference,liu}


\end{document}